\documentclass[iop]{emulateapj}
\usepackage{epsfig}
\usepackage{amsmath}
\usepackage{graphicx}
\usepackage{natbib}

\begin{document}

\title{On the Lack of a Radio Afterglow from Some Gamma-Ray Bursts - Insight into Their Progenitors?}
\author{Nicole M. Lloyd-Ronning$^{1,2}$ \& Christopher L. Fryer$^{1,2}$}
\affil{$^1$ CCS-2, Los Alamos National Lab, Los Alamos, NM 87544 \\
$^2$ Center for Theoretical Astrophysics, Los Alamos National Lab, Los Alamos, NM 87544}

\begin{abstract} 
  We investigate the intrinsic properties of a sample of bright (with isotropic equivalent energy $E_{iso}>10^{52}$ erg) gamma-ray bursts, comparing those with and without radio afterglow.  We find that the sample of bursts with no radio afterglows has a significantly shorter mean intrinsic duration of the prompt gamma-ray radiation, and the distribution of this duration is significantly different from those bursts with a radio afterglow.  Although the sample with no radio afterglow has on average lower isotropic energy, the lack of radio afterglow does not appear to be a result of simply energetics of the burst, but a reflection of a separate physical phenomenon likely related to the circumburst density profile.  We also find a weak correlation between the isotropic $\gamma-$ray energy and intrinsic duration in the sample with no radio afterglow, but not in the sample which have observed radio afterglows.  We give possible explanations for why there may exist a sample of GRBs with no radio afterglow depending on whether the radio emission comes from the forward or reverse shock, and why these bursts appear to have intrinsically shorter prompt emission durations.  We discuss how our results may have implications for progenitor models of GRBs.
\end{abstract}

\keywords{stars: gamma-ray bursts: general}

\section{Introduction}
   Over the last couple of decades, we have gained a general understanding of the nature of gamma-ray bursts (GRBs) thanks to satellites such as BATSE, BeppoSAX, Swift, Fermi, Integral, and many others, not to mention the many ground-based follow-up observations of GRB afterglows (for a summary of results, see reviews by van Paradijs et al. 2000; Piran 2004; Zhang \& Meszaros 2004; Meszaros 2006; Woosley \& Bloom 2006; Gehrels, Ramirez-Ruiz \& Fox 2009; Gombec 2012; Berger 2014).  However, these data have also opened up many new questions, and in some ways increased the number of viable models for GRBs.  Although it appears likely GRBs are associated with a massive stellar progenitor in the case of long bursts (see, e.g., Gehrels, Ramirez-Ruiz \& Fox 2009 and references therein for a summary of the observations; for an early theoretical perspective see Popham, Woosley, \& Fryer 1999), and a merger of compact objects in the case of short bursts \cite{Berg14}, uncertainties in the details of the progenitor remain.  

   It is clear that in order to shed light on the nature of GRBs, we must continue to analyze their broad-band spectral data and light curves. In particular, features in the light curves such as plateaus, flares, and steep decays (Swenson et al. 2013; Swenson \& Roming 2014) all contain clues to the dissipation mechanism and, potentially, the progenitor. One of the most difficult aspects of GRBs, however, is the number of parameters that can play a role in the behaviors of the spectra and light curves.  In addition to global burst parameters (such as the energy emitted, the duration of the inner engine, the density profile of the circumburst medium), the microphysical parameters (e.g. fraction of energy in the magnetic field, and the energy distribution of radiating particles) can significantly affect the resultant light curve.  The degeneracy among all of the various physical parameters make it difficult to draw firm conclusions about the detailed physics of GRBs.

 In this paper, we examine the properties of GRBs with and without radio afterglows, in an attempt to gain insight into the inner engine and environment of a GRB. Because the radio afterglow generally peaks at later times and the emission mechanism is fairly solidly understood as synchrotron emission, we can circumvent some uncertainties that arise in the X-ray and optical afterglow emission (Frail et al. 1997; Galama et al. 2000). The long-lived radio afterglow may be a better probe (compared to optical and X-ray) of the far-out circumstellar environment of the GRB, and offer a better estimate of the energetics (see, e.g., Frail, Waxman, \& Kulkarni 2000) which can, in turn, help us learn something about the progenitor.
  
   Chandra \& Frail (2012) presented analysis of a sample of 304 GRBs that were followed up in the radio over a span of 14 years, from 1997-2011.  They carried out a number of statistical analyses, and found a detection rate of radio afterglows around $31\%$.  This is in sharp contrast to the detection rates of X-ray afterglows ($\sim 95 \%$) and optical afterglows ($\sim 70 \%$).  The radio light curve at 8.4GHz peaks at around 3-6 days with a median peak luminosity of $10^{31}$erg s$^{-1}$ Hz$^{-1}$. Although they suggest there is a relationship between the detectability of a radio afterglow and the fluence or energy of the GRB, they find no significant correlations between the strength of the radio flux density and the GRB energy, fluence or X-ray flux (they do find a mild correlation between the strength of the optical flux density at 11hr and the peak radio flux density; see their Table 5). Ultimately, they conclude that radio afterglow samples are sensitivity-limited and therefore bursts without radio afterglow are not inherently radio quiet.

   However, Hancock, Gaensler, \& Murphy (2013) - hereafter HGM - present an alternative view.  By using visibility stacking techniques (see the method described in Hancock, Gaensler \& Murphy 2011) of 737 radio observations consisting of 178 GRBs for which VLA data could be calibrated, they showed the stacked data of radio faint GRBs did not produce any detections.  Instead, they suggest that there are two intrinsically different populations of GRBs - radio loud and radio quiet (interestingly, after decades of debate, it has emerged that there is a bimodal distribution of quasars in terms of the presence/absence of radio emission, and this appears to be a true, physical - i.e. not a selection - effect; see Kellerman et al. 2016 and references therein). They estimate that $\sim 30-40 \%$ of GRBs are truly intrinsically radio quiet.  Although they find that the redshift distributions between radio loud and faint are statistically the same, they claim signficant differences between their radio loud and quiet samples in observed prompt duration, gamma-ray fluence, optical and x-ray flux, and isotropic equivalent energy (see their Table 3). They speculate that the inherent difference between radio loud and faint GRBs could be a reflection of either different emission mechanisms, or a magnetar-driven engine (radio quiet) vs. black-hole driven engine (radio loud).  Indeed bimodality in GRB afterglow emission - particularly in the temporal decay indices and luminosities - has previously been suggested in X-rays (Boer \& Gendre 2000; Gendre \& Boer 2005; Gendre et al. 2008) and optical (Nardini et al. 2006; Liang \& Zhang 2006).
 
   Motivated by HGM, we investigate further the suggestion of an intrinsically radio quiet sub-population of GRBs. In particular, we examine a sample of the intrinsically brightest GRBs (in terms of isotropic emitted $\gamma-$ray energy $E_{iso}$).  Selecting only those bursts with $E_{iso} > 10^{52} erg$, we investigate the differences of various GRB {\em intrinsic} properties, for those bursts with and without a radio afterglow. We find a significant difference in the mean isotropic $\gamma-$ ray energy between the radio loud and quiet samples (similar to Chandra \& Frail 2012 and HGM, who made no cut in $E_{iso}$).  Interestingly, however, we also find a significant difference between the intrinsic duration distributions.  We find the latter difference to be the most robust when comparing the radio loud and radio quiet samples.  In addition, we find a weak correlation between isotropic energy and intrinsic duration in the radio quiet sample, but not the radio loud sample.  We also compare radio loud and quiet samples for intrinsically faint GRBs (in terms of isotropic energy) and find no difference among the two samples.  We explore the implications our findings have on progenitor models.  
 
  This paper is organized as follows: In \S 2, we describe our data sample taken from the Chandra \& Frail (2012) catalog.  In \S 3, we present our statistical analysis of the bright bursts ($E_{iso} > 10^{52}$ erg) with and without a radio afterglow. For comparison, we also present an analysis of radio loud and quiet bursts with energies below our energy cutoff. In \S 4, we review some of the physical parameters that play a role in determining the strength of the radio flux from the forward and reverse shock of the external blast wave.  In \S 5, we discuss the implications of our results for progenitor models of GRBs.  Conclusions are presented in \S 6.  Throughout the paper we use the term ``radio loud''  and ``radio quiet'' to refer to GRBs with and without a detected radio afterglow, respectively. 

\section{Data Sample}
   We begin with the catalog from Chandra \& Frail (2012) - a sample of 304 GRBs for which radio follow-up observations were performed. We choose those bursts with redshift measurements so we can determine intrinsic properties such as energy emitted and intrinsic duration.  Note that Turpin et al. (2016) examined possible biases in selecting bursts with measured redshifts; they found little bias in the properties of the bursts, {\em except} for a mild preference (at a significance level $\sim 2.5 \sigma$) for those bursts with measured redshifts toward slightly higher observed prompt duration $T_{90}$.  Because we are comparing samples of bursts within the subset of GRBs that have redshifts and because the bias toward longer duration is weak, we do not expect this to have a significant impact on our results.
 We then select the bursts which are intrinsically bright in terms of isotropic $\gamma-$ray emitted energy, $E_{iso} > 10^{52} erg$. This leaves us with a total sample of 96 bursts.  Of these bursts, 59 have a radio afterglow and 37 have no radio afterglow. The data we used in this analysis is given in Tables 7 and 8 at the end of this paper.

\subsection{Selecting Bursts with $E_{iso} > 10^{52}$ ergs}
 One can ask whether we should use the actual (beaming-corrected) energy emitted from the GRB instead of the isotropic-equivalent energy (with no correction for the beamed jet).  Unfortunately, jet break measurements are tenous (not to mention highly uncertain), and restricting ourselves to those with jet breaks would prohibitively limit our sample and introduce additional unknown biases.  However, it is a reasonable assumption that the isotropic equivalent energy is a fair estimate of the true energy of the GRB.  Figure 1 shows $E_{iso}$ vs. the beaming-corrected energy $E_{jet}$ taken from Ghirlanda et al. (2004).  A Kendell’s $\tau$ test on this $E_{iso}$ vs. $E_{jet}$ sample gives a $> 4 \sigma$ correlation between the two values.  Hence, we use $E_{iso}$ as a reasonable proxy for the true emitted energy of the GRB.

\begin{figure}
\begin{center}
\epsfxsize=9cm\epsfbox{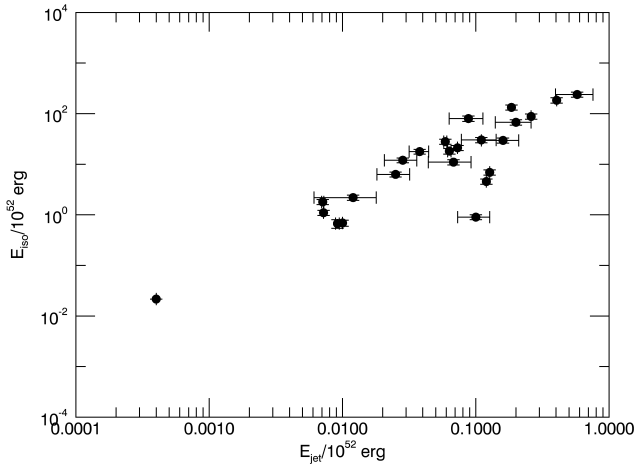}
\caption{Isotropic equivalent energy $E_{iso}$ versus energy corrected for the opening angle of the jet $E_{jet}$.  Data is taken from Ghirlanda et al. 2004.}
\end{center}
     \label{}
\end{figure}

 Our goal in this paper is to try to identify and examine a sample of GRBs that may be intrinsically radio quiet, and compare it to a sample of bursts which have radio afterglows.  However, as discussed in HGM, the sample of bursts with no radio afterglow detection is contaminated by bursts with  radio fluxes that simply fell below detection sensitivity limits (and are not truly radio quiet).  Recall that Chandra \& Frail (2012) found that brighter bursts (higher $E_{iso}$) tend to have greater detectability of their radio afterglows (although there was no direct correlation between the strength of the radio flux and $E_{iso}$; see their Table 5).  Therefore, our cut in $E_{iso}$ is our attempt to pick a sample least contaminated by faint radio afterglows that have fallen below detector threshholds, and define a truly radio quiet sample.  Nonetheless, even if our $E_{iso}$ cut sample is not an unbiased way to examine the radio loud and quiet bursts, we can at the very least ask the question of whether there is a difference between bursts of high $E_{iso}$ with and without radio afterglows. We present this analysis below.


\section{Two Populations}
  We investigate whether there are any inherent differences between those bursts with and without a radio afterglow for our sample of intrinsically bright bursts (isotropic equivalent gamma-ray energy, $E_{iso} > 10^{52}$ erg). Table 1 presents the mean values of redshift $z$, intrinsic duration of the prompt emission $T_{int} = T_{90}/(1+z)$, and isotropic equivalent energy $E_{iso}$ for the radio loud and quiet bursts in our sample.  A Student's t-test (Student 1908) on the radio loud and quiet samples gives a probability of $0.004$ that the average intrinsic durations are consistent between the two samples, and a probability of $0.002$ that the averages of $E_{iso}$ are consistent (a t-test between the means of the redshifts did not yield a signficant difference).

 The data suggest that the values of duration and energy in particular are indicative of a difference between the two populations.  To quantify this statement, we perform Kolmogorov-Smirnov (KS) tests (Kolmogorov 1933) on the cumulative distributions of redshift, isotropic energy and intrinsic duration, comparing the radio loud and quiet bursts.  
Table 2 presents the KS test probabilities (the probability that the samples are drawn from the same parent distribution), comparing the radio loud and quiet samples for the set of bursts with $E_{iso} > 10^{52} erg$.  The distributions of intrinsic duration and isotropic energy show a significant difference, while the redshift distribution does not (we consider a KS probability $< 10^{-3} \sim 3 \sigma$ signficant).  We also compared the samples over the same duration and energy range, and found that the intrinsic duration distributions remain significantly different, but the energy distribution does not. Figures 2 and 3 show the cumulative intrinsic duration and isotropic energy distributions of our sample.  For reference, we have included the sample of {\em all} (with no isotropic energy cutoff) Swift GRBs with radio afterglows (dotted line).

Table 3 shows the KS tests for the sample of radio loud and quiet bursts with $E_{iso}<10^{52} erg$.  {\em There is no significant difference between the cumulative distributions of any of the three intrinsic properties we have analyzed for these lower energy bursts.}  We attribute this to the radio faint sample being contaminated by bursts that are not truly intrinsically radio quiet, but whose radio afterglows simply fell below the detector sensitivity limit and therefore are largely from the same population as the radio loud bursts.

\begin{deluxetable}{lccc}
\tablecaption{Average Values of Intrinsic Properties}
\tablecolumns{4}
\tablewidth{\linewidth}
\tablehead{Sample & $\bar{z}$ & $\bar{T}_{int}$ (s) & $\bar{E}_{iso}(10^{52}$ erg)}
\startdata
 Bright, No Radio (37 bursts) | & 2.7$\pm 0.3$ & 16.$\pm 3.$ & 9.$\pm 2.$ \\ 
 Bright, Radio  (59 bursts) | & 2.1$\pm0.2$ & 38.$\pm 6.$ & 50.$\pm 10.$ \\ \\
 Dim, No Radio (29 bursts) | & .82$\pm 0.1$ & 34.$\pm 11.$ & .28 $\pm 0.1$\\ 
 Dim, Radio (14 bursts) | & .55 $\pm  0.1$ & 35.$\pm 16.$ & .35 $\pm 0.1$
\enddata
\tablecomments{   Average values of the redshift $z$, intrinsic prompt duration $T_{int}$, and isotropic emitted energy $E_{iso}$ for the sample of bright ($E_{iso} > 10^{52}$ erg) and dim ($E_{iso}< 10^{52}$ erg) bursts with and without a detected radio afterglow. \\  }
\label{}
\end{deluxetable}


\begin{deluxetable}{lc}
\tablecaption{KS Tests Between Samples - Bright ($E_{iso} > 10^{52} erg$) Bursts}
\tablecolumns{4}
\tablewidth{\linewidth}
\tablehead{ Property & KS Probability}
\startdata
 Intrinsic Duration (entire samples) | & $3\times 10^{-4}$ \\
 Isotropic Energy (entire samples) | & $1\times 10^{-4}$ \\
 Redshift | & $.05$ \\
 Intrinsic Duration ($1-80$ s) | & $1\times 10^{-3}$ \\
 Isotropic Energy ($1-40 \times 10^{52}$ erg) | & .02 
\enddata
\tablecomments{Comparison of the cumulative distributions of intrinsic duration, isotropic equivalent energy and redshift between the sample with radio afterglows and the sample without, for so-called bright bursts with $E_{iso} > 10^{52}$ erg.  The Kolmogorov-Smirnov (KS) probability values give the probability that the samples are drawn from the same distribution.  Both intrinsic duration and isotropic energy show significantly different distributions when comparing the entire samples.  When we compare the distributions over the same ranges of duration/energy, we find only the duration distribution remains statistically significantly different.  The redshift distributions are not statistically significantly different.}
\label{}
\end{deluxetable}

\begin{deluxetable}{lc}
\tablecaption{KS Tests Between Samples - Dim ($E_{iso} < 10^{52} erg$) Bursts}
\tablecolumns{4}
\tablewidth{0.85\linewidth}
\tablehead{ Property & KS Probability}
\startdata
 Intrinsic Duration | & $.57$ \\
 Isotropic Energy | & $.80$ \\
 Redshift | & $.3$ 
\enddata
\tablecomments{Comparison of the cumulative distributions of intrinsic duration, isotropic equivalent energy and redshift between the sample with radio afterglows and the sample without, for so-called dim bursts with $E_{iso} < 10^{52}$ erg.  The Kolmogorov-Smirnov (KS) probability values give the probability that the samples are drawn from the same distribution. None of the distributions show a signficant difference.\\ } 
\label{}
\end{deluxetable}

  Our results suggest that there is something inherently different about $\gamma-$ray bright bursts with no radio afterglow.  We acknowledge we are dealing with small number statistics and small biases can skew results.  Our choice of comparing only those bursts with high $E_{iso}$ is an attempt to minimize these biases, and best isolate a sample of truly radio quiet bursts. However, acknowledging the biases that come into play when making cuts on the data, we also tried a different cut - in X-ray flux at 11hr.  Taking those bursts with $f_{X,11hr} > 10^{-12} erg s^{-1} cm^{-2}$, we found no difference between the radio loud and radio faint populations' intrinsic properties (duration, isotropic energy and redshift), with KS probabilities of $0.6$ (intrinsic duration), $0.9$ (isotropic energy), and $0.3$ (redshift).  Hence, we conclude that it is the isotropic equivalent energy cut that is allowing us to better distinguish potential inherent differences between radio loud and faint populations.

\begin{figure*}[htbp]
\centering
\includegraphics[width=5.0in]{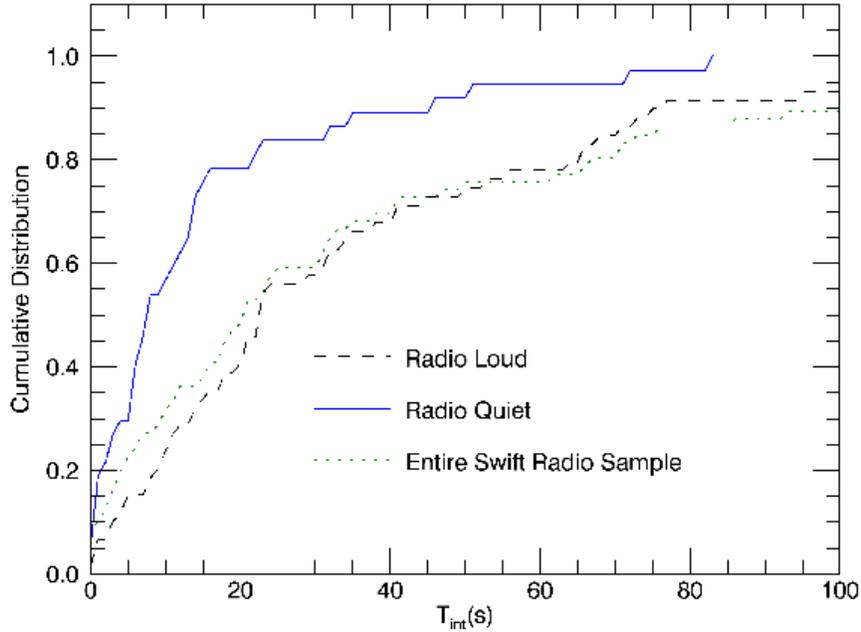}
\caption{Intrinsic duration $T_{int}$ for the Chandra \& Frail 2012 bright ($E_{iso}>10^{52}$ erg) burst sample with radio afterglows (dashed black line) and without (solid blue line). For reference we have also plotted the entire Swift sample with radio afterglow detections, and with no energy threshold cutoff (dotted green line). }
\end{figure*}

\begin{figure}
\begin{center}
\epsfxsize=9cm\epsfbox{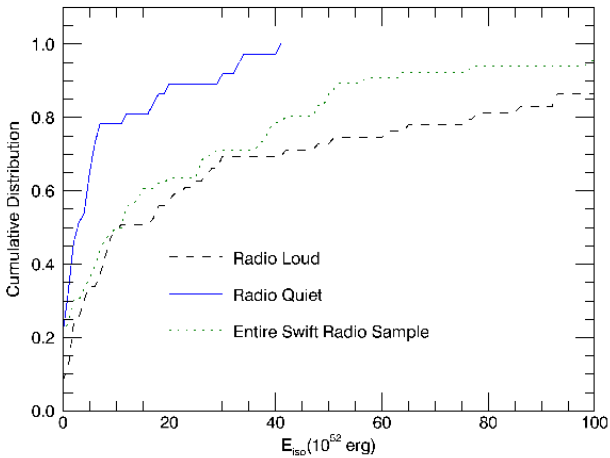}
\caption{Isotropic equivalent energy $E_{iso}$ for the Chandra \& Frail 2012 bright ($E_{iso}>10^{52}$ erg) burst sample with radio afterglows (dashed black line) and without (solid blue line). For reference we have also plotted the entire Swift sample with radio afterglow detections, and with no energy threshold cutoff (dotted green line).}
\end{center}
     \label{}
\end{figure}

\subsection{Correlation Analysis}
Figure 4 shows $E_{iso}$ vs. intrinsic prompt duration $T_{int}$ for the radio quiet and loud samples. The horizontal dotted line marks our $E_{iso}$ cutoff.  In the left panel for the radio quiet sample, we mark a shaded region where a potential observational bias may be playing a role.  Although our sample has no simple single flux limit for burst detection (see swift.gsfc.nasa.gov for a detailed description of GRB detection criteria and detector response matrices), in general bursts with a given $E_{iso}$ but longer duration will have a lower flux, potentially putting these bursts below the detection threshold. This selection effect could serve to artificially enhance or produce a correlation when none is there.  Ideally, we would like to address this issue using non-parametric statistical techniques designed to handle observational truncations in the data and reproduce true underlying correlations (Lynden-Bell, 1971; for application to GRBs see Lee \& Petrosian, 1996 and Lloyd et al., 2000).  However, to do so, we need both a larger sample and a better handle on the detection criteria for each burst in our sample.

 Keeping in mind selection effects may be artificially producing the $E_{iso}-T_{int}$ correlation, we nonetheless employ a Kendell\'s $\tau$ test (Kendell 1938) on our samples.  We find a mild correlation ($\sim 3 \sigma$) between the duration and isotropic energy for bursts without radio afterglows.  Parameterizing this correlation, we find $E_{iso} \propto T_{int}^{0.44 \pm 0.15}$.  We discuss the implication of this potential correlation on the progenitors of GRBs below. For radio loud bursts, we found no correlation between the isotropic emitted energy and intrinsic duration (or any other observed or intrinsic property for which the data was available).

\begin{figure*}[htbp]
\centering
\includegraphics[width=3.5in]{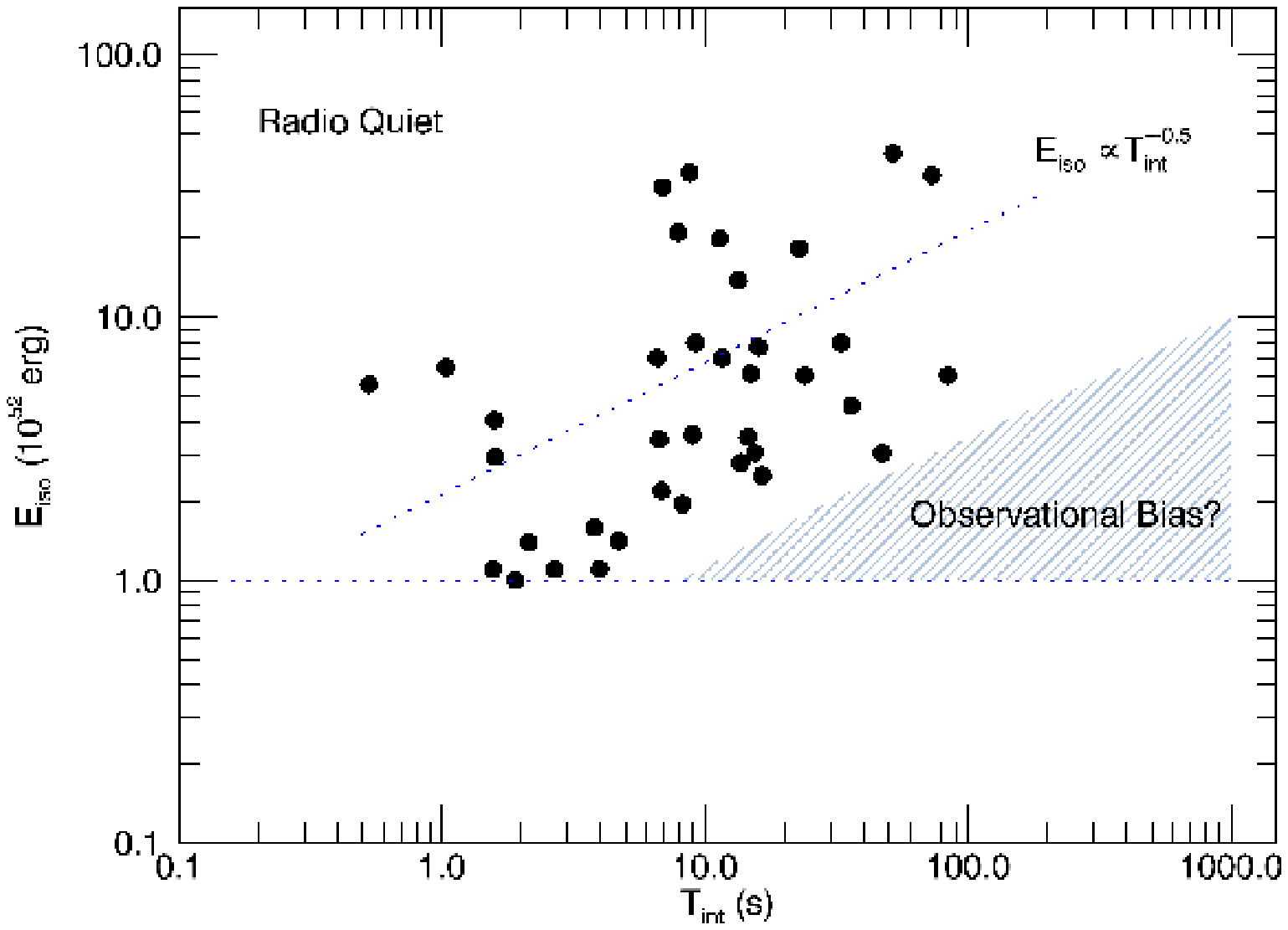}\includegraphics[width=3.5in]{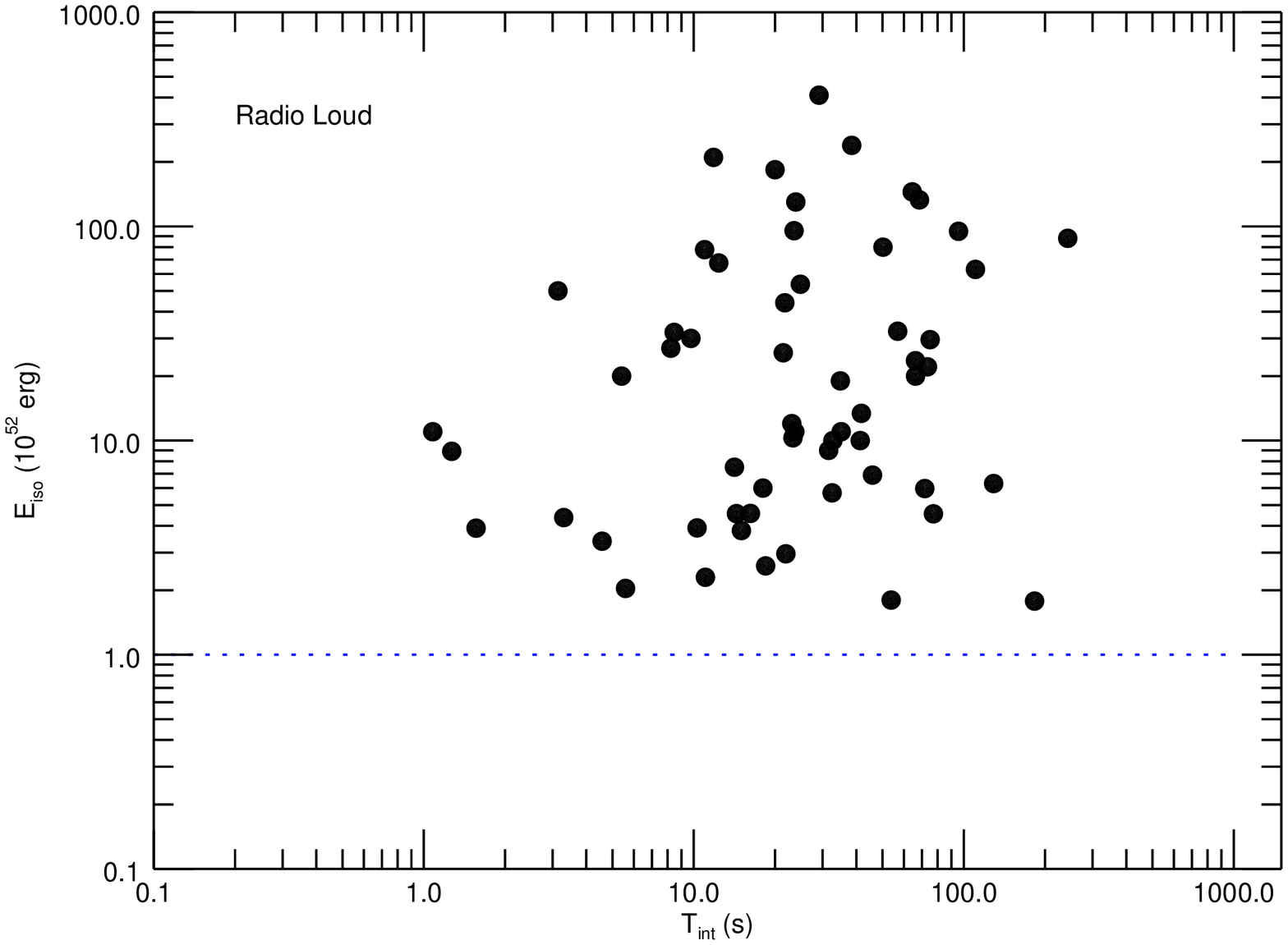}
\caption{{\em Left panel:} Isotropic equivalent energy $E_{iso}$ vs. intrinsic duration $T^{\prime}_{int}$ for the sample of bright bursts with no radio afterglow. A Kendell's $\tau$ test indicates a weak correlation ($\sim 3 \sigma$) between the two, where $E_{iso} \sim (T^{\prime}_{int})^{1/2}$ The shaded region marks potential observational bias for a fixed flux limit. {\em Right panel:} Isotropic equivalent energy $E_{iso}$ vs. intrinsic duration $T^{\prime}_{int}$ for the sample of bright bursts with a radio afterglow. A Kendell's $\tau$ test indicates {\em no correlation} between the two variables. The horizontal dotted line marks the energy cut for our samples.}
\end{figure*}

\section{Radio Afterglow Flux}
In interpreting our results above, we would like to get an understanding of the role various physical parameters play in determining the observed brightness of the radio afterglow flux.  A GRB afterglow is most commonly modeled as coming from synchrotron emission from the outflow shocking with the external medium (Meszaros \& Rees, 1997; Sari, Piran, \& Narayan 1998).  Both a forward and reverse shock can contribute to the emission to various degrees, and although most studies have modeled the emission as coming from a {\em forward} shock (e.g. Frail et al. 2000, Yost et al. 2003), there have been suggestions that the radio afterglow (especially at early times $< 10$ days) is a result of the {\em reverse} shock emission (Laskar et al. 2013, Laskar et al. 2016).
Many papers have estimated the GRB emission from the forward and reverse shock (e.g. Meszaros \& Rees 1997; Sari, Piran \& Narayan 1998; Chevalier \& Li 2000, Kobayashi 2000, Granot \& Sari 2002, Zou et al. 2005).  Here we summarize some main results, using the formalism adopted in Granot \& Sari (2002).

\subsection{Forward Shock}
The presence of synchrotron emission in the radio band is expected from the interaction of the forward shock with the external medium.
In what follows, we define the parameter $p$ as the index of the electron energy distribution (where electrons are distributed as a power law in energy, $dN/d\gamma_{e} \propto \gamma_{e}^{-p}$), $\epsilon_{B}$ and $\epsilon_{e}$ are the fractions of energy in the magnetic field and electrons respectively, $n$ is the ISM density normalized to $cm^{-3}$, $A_{\star}$ is the normalization of the density distribution in a wind medium ($\rho = 5\times 10^{11} g cm^{-1} A_{\star}r^{-2}$, where $r$ is the radius), $E_{52}$ is the isotropic equivalent energy normalized to $10^{52}$ erg, and $d_{L,28}$ is the luminosity distance normalized to $10^{28}$ cm.

In the standard external shock model, the observed flux will peak at one of three break frequencies - the so-called cooling frequency $\nu_{c}$ which defines a characteristic energy above which electrons rapidly lose all of their energy to radiation; the minimum electron frequency $\nu_{m}$, which is defined by the minimum energy or break in the power-law distribution of electron energies; and the self-absorption frequency $\nu_{a}$, below which synchrotron photons are absorbed and the flux is correspondingly suppressed (see, e.g., Granot \& Sari 2002 for details).  The cooling frequency is expected to be well above the radio band for fiducial GRB parameters and relies on the simplistic assumption that particles are instantaneously accelerated and then subsequently radiate all of their energy (more likely, acceleration processes are continual as suggested in Lloyd \& Petrosian 2000, Lloyd-Ronning \& Petrosian 2002).  We do not consider the cooling frequency here.  Therefore, the peak flux of the radio emission will occur either at $\nu_{m}$ or $\nu_{a}$.

For $\nu_{a} < \nu_{m}$, the peak flux in an ISM medium is:
\begin{equation}
\begin{aligned}
f_{p}(\nu_{m}) = & 9.93mJy(p+0.14)(1+z) \\ 
& \epsilon_{B}^{1/2}n_{cm^{-3}}^{1/2}E_{52}d_{L,28}^{-2}
\end{aligned}
\end{equation}

For a wind medium we have:
\begin{equation}
\begin{aligned}
f_{p}(\nu_{m}) = & 76.9mJy(p+0.12)(1+z)^{3/2} \\
& \epsilon_{B}^{1/2}A_{\star}E^{1/2}_{52} t^{-1/2}_{days} d_{L,28}^{-2}
\end{aligned}
\end{equation}

For $\nu_{m} < \nu_{a}$ (which may occur, for example, in a high density medium or for low values of the electron energy fraction; see Table 2 of Granot \& Sari 2002 for these dependencies), the peak flux is at the self-absorption frequency and we have for an ISM medium:
\begin{equation}
\begin{aligned}
f_{p}(\nu_{a}) =& 20.8mJy(p-1.53)e^{2.56p} d_{L,28}^{-2} \\
&\left(\frac{(1+z)^{7p+3}\epsilon_{B}^{2p+3}E_{52}^{3p+7}}{\epsilon_{e}^{10p-10}t^{5(p-1)}}\right)^{\frac{1}{2(p+4)}}
\end{aligned}
\end{equation}

For a wind medium, we have:
\begin{equation}
\begin{aligned}
f_{p}(\nu_{a}) = & 158.mJy(p-1.48)e^{2.24p} d_{L,28}^{-2} \\
&\left(\frac{(1+z)^{6p+9}\epsilon_{B}^{2p+3}E_{52}^{4p+1}}{\epsilon_{e}^{10p-10}A_{\star}^{2(p-6)}t^{4(p+1)} }\right)^{\frac{1}{2(p+4)}}
\end{aligned}
\end{equation}

 It is obvious how highly degenerate the spectra and light curves are (i.e. how many combinations of the different parameters can produce the same results).  The ordering of the characteristic frequencies and values of the peak fluxes are sensitive not only to the blast wave energy and external density, but can also be highly sensitive to the microphysical parameters, such as the fraction of the energy in the radiating electrons and magnetic field.  Nonetheless, for fiducial GRB parameters of the forward shock, the light curve in a wind medium will in general decay {\em faster} than in a constant density ISM-like medium. Intriguingly, we found that the X-ray luminosity at 1 hour and 11 hours was slightly higher in the radio quiet sample than radio loud, potentially suggestive of a more quickly decaying light curve for the radio quiet sample (we point out, however, the numbers here were very small - only $\sim 10$ bursts in each sample had this data available).

\subsection{Reverse Shock}
 Emission is also expected as the reverse shock travels back through the the ejecta (e.g. Meszaros \& Rees 1997, Sari \& Piran 1999, Kobayashi 2000).  Initially it was suggested that the reverse shock would produce short lived optical emission, and indeed this is the interpretation for a few observed optical flashes (Sari \& Piran 1999, Vestrand et al. 2014).  Others have suggested the reverse shock emission is more readily observed at longer wavelengths (Mundell et al. 2007, Laskar et al. 2013, Kopac et al. 2015).  Indeed Laskar et al. 2013 and Laskar et al. 2016 show that the early time radio emission in GRB 130427A and 160509A is best fit by a Newtonian reverse shock in a low density medium ($n \sim 10^{-3} cm^{-3}$).  

  As with the forward shock, there are many parameters that come into play in determining the energy range and strength of the peak flux of the reverse shock (see, e.g., Kobayashi 2000 and Zou et al. 2005 for expressions for the break energies and peak fluxes of the reverse shock in an ISM and wind medium respectively).  The thickness of the shell through which the reverse shock travels, the microphysical parameters, and the external density profile all play a role in determining the strength and spectral energy band of the reverse shock emission. Roughly, the ratio of peak flux of the reverse shock to the forward shock at $\nu_{m}$ is given by (Laskar et al. 2016):
$f_{\nu_{m}, RS}/f_{\nu_{m}, FS} \sim \Gamma (\epsilon_{B, RS}/\epsilon_{B, FS})^{1/2}$.  However, the frequency of the emission peaks $1/\Gamma^{2} (\epsilon_{B, RS}/\epsilon_{B, FS})^{1/2}$ lower than the forward shock.  Note that if the cooling frequency $\nu_{c}$ plays a role in the emission (although, again, it may not be relevant depending on the acceleration processes) and falls in the radio band the reverse shock emission vanishes above this frequency (see equation 14 of Kobayashi 2000).

  In addition, if the radio afterglow is from the reverse shock, the emission needs to last sufficiently long to accommodate the radio afterglow observations.  Because the reverse shock travels back into the relativistic ejecta and its duration is determined by the time it takes to cross this ejecta, this implies that the deceleration radius of the outflow $R_{dec}$ (which marks the on-set of the afterglow; Meszaros \& Rees 1999, Sari, Piran, \& Narayan, 1999) be sufficiently far out (or alternatively, that the afterglow begins at sufficiently late times). However the deceleration radius differs between a wind and constant ISM-like circumburst density, and occurs generally much closer in (or at earlier times) for a wind medium. For an ISM-like medium, we have
\begin{equation}
R_{dec,ISM} \approx 2\times10^{16}cm (E_{52})^{1/3}(n_{1})^{-1/3}(\Gamma_{400})^{-2/3}
\end{equation}

\noindent For a wind medium, we have
\begin{equation}
R_{dec,wind} \approx 10^{13}cm (E_{52})(A_{\star})^{-1}(\Gamma_{400})^{-2}
\end{equation}
 
\noindent where the deceleration timescale is given by $t_{dec} = R_{dec}/c\Gamma^{2}$. If indeed the presence of the radio afterglow reflects the presence of emission from the reverse shock, it is possible that the reverse shock emission in a wind medium is too short lived to be detected. This has important implications for the progenitor as discussed below.

\subsection{Energy and redshift factors}
  Although we found no significant difference between the cumulative redshift distributions of our radio loud and quiet samples (and recall their average values differ by a factor of only $\sim 1.4$), we can ask to what extent redshift and energy play a role in the detection of the radio flux. It is well established that GRBs are not standard candles in energy (although a narrow distribution of beaming-corrected energies was found by Frail et al. 2001, this distribution was later shown to be much broader than originally thought. See, e.g. Kocevski \& Butler 2008 and Li 2008). Hence, redshift (distance) alone is unlikely to play a role in the lack of radio detection, and indeed we find no significant (anti-) correlation between radio flux and redshift.  And although the population with no radio afterglow has an average lower isotropic equivalent energy (by a factor of $\sim 5.5$), there is also no correlation between $E_{iso}$ and radio flux (as born out by a Kendell's $\tau$ test or Spearman rank order test; see also Figure 5), so we cannot simply conclude the lower isotropic energy is the reason for lack of detection. In any case, there is not an obvious reason why those intrinsically bright (in terms of $E_{iso}$) bursts with no radio afterglow should have a significantly different intrinsic duration distribution than the bursts with radio afterglows.  
  

If we compare our radio loud and quiet samples over a redshift range such that the samples have the same mean redshift ( $\bar{z} \approx 2.0$), we find that our primary result holds - the duration distributions remain significantly different, with radio quiet bursts being on average shorter ($\bar{T}_{int} = 17. \pm 4.$) than radio loud ($\bar{T}_{int} = 39. \pm 6.$).  Similarly, when we compare the radio quiet and loud samples over the same energy range (that spanned by the radio quiet sample, which had a smaller range of $E_{iso}$), we find that the duration distributions remain significantly different, with the radio quiet sample on average shorter ($\bar{T}_{int} = 16. \pm 3.$) compared to the radio bright ($\bar{T}_{int} = 35. \pm 6.$). Table 4 shows the results of a KS test analysis on the duration distributions of these radio bright and dark sub-samples, indicating the difference between their duration distributions.

\begin{deluxetable}{lc}
\tablecaption{KS Tests, Radio Dark and Bright sub-samples}
\tablecolumns{2}
\tablewidth{\linewidth}
\tablehead{ Property & KS Probability}
\startdata
 Intrinsic Duration (same mean redshift) | & $7\times 10^{-4}$ \\
 Intrinsic Duration (same energy range) | & $2\times 10^{-3}$ 
\enddata
\tablecomments{Comparison of the cumulative distributions of intrinsic duration for radio dark and bright samples for two cuts in the data. The first cut compares the durations of the radio bright and dark samples such that the samples have the same mean redshift.  The second compares the durations of theradio bright and dark samples over the same energy range (that spanned by the radio dark sample).}
\label{}
\end{deluxetable}

 
  Finally, we can input the observed mean redshift and energy of our radio loud and quiet samples from Table 1 into equations 1-4, and compute the fluxes.  In this case, we find the flux in the radio bright sample is higher by a factor of $\sim 3.5 -8$ depending on whether the medium is ISM or wind and whether the peak of the spectrum occurs at $\nu_{m}$ or $\nu_{a}$.  Meanwhile, observations span at least two orders of magnitude. Table 5 summarizes these results (note that we have assumed constant values for $p$, $\epsilon_{B}, \epsilon_{e}, A_{*}$).  


\begin{deluxetable}{lc}
\tablecaption{Ratio of Expected Peak Radio Fluxes}
\tablecolumns{2}
\tablewidth{\linewidth}
\tablehead{Model & Ratio of Peak Flux}
\startdata
 $f_{\nu_{m}, ISM}(R)/f_{\nu_{m}, ISM}(NR)$  & $8.0$ \\
 $f_{\nu_{m}, wind}(R)/f_{\nu_{m}, wind}(NR)$  & $3.5$ \\
 $f_{\nu_{a}, ISM}(R)/f_{\nu_{a}, ISM}(NR)$ & $7.5$ \\
 $f_{\nu_{a}, wind}(R)/f_{\nu_{a}, wind}(NR)$ & $5.0$
\enddata
\tablecomments{Ratio of the radio peak fluxes expected for given the mean of the isotropic energy and redshift of the radio to no-radio sample,where R denotes the radio sample with mean $\bar{E}_{iso} = 50 \times 10^{52}$ erg, $\bar{z}=2.0$, while NR denotes the radio quiet sample with mean $\bar{E}_{iso} = 9 \times 10^{52}$ erg, $\bar{z}=2.6$.  The ratio is calculated using equations 1-4 above, in which the peak occurs at $\nu_{m}$ or $\nu_{a}$ for either an ISM or wind circumburst medium.} 
\label{}
\end{deluxetable}

\begin{figure}
\begin{center}
\epsfxsize=9cm\epsfbox{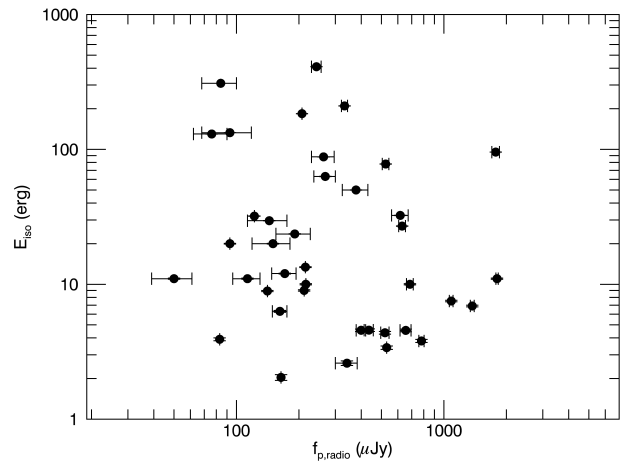}
\caption{Peak radio flux versus isotropic equivalent energy for the radio loud sample. There is no correlation between the two quantities.}
\end{center}
     \label{}
\end{figure}

\section{Implications for Progenitors}
  Our main result is that there is a significant difference in the intrinsic prompt duration of (bright) bursts with and without a radio afterglow, where radio quiet bursts tend to have on average a shorter duration (and a significantly different cumulative distribution even when comparing over the same duration ranges). Similar to others - albeit analyzing different samples - we find that the sample with radio afterglows has a higher $E_{iso}$ on average and a different cumulative distribution of $E_{iso}$, although the difference is minimized when we compare over the same range of energies.  Additionally, we find a mild correlation between intrinsic duration and isotropic $\gamma-$ray energy for bursts without a radio afterglow (but not those with a radio afterglow), although observational biases may be playing a role in producing this correlation.  

It is worthwhile to consider these results in the context of various progenitor models of GRBs.  In general, the observables are connected to the progenitor in ways that are not always straightforward and - as mentioned in \S 4 -  many factors can come into play. However, we can look at general trends and expectations from different models. Table 6 gives a summary of important variables for different GRB progenitors, which can be connected to the relevant observables.

\begin{deluxetable}{lcccr}
\tablecaption{Features of GRB Progenitor Models}
\tablecolumns{5}
\tablehead{Model & Ang. Mom. & Acc. Rates & Fuel & Mass loss}
\startdata
 Collapsar & Low & High & High & High \\ 
 Helium Merger  & High & Low-High & Moderate & Moderate  \\ 
 WD-NS/BH  & High & Low & Limited & Low/ISM \\ 
 NS-NS/BH  & Limited & High & Limited & Low/ISM \\
 Magnetar & Limited & Low/None & Limited & Low-High  
\enddata
\tablecomments{A summary of expected properties of different GRB progenitor models.  See \S 5.1 for a more detailed discussion of each model.  \\  }
\label{}
\end{deluxetable}

\subsection{GRB Progenitor Models}
Below we discuss the general properties (particularly energetics and duration) of some of the most common models for GRB progenitors.
\begin{itemize}
 \item{ {\bf Collapsar:} In the most generic version of the collapsar model of GRBs (see, e.g., Woosley 1993; MacFadyen \& Woosley 1999), a very massive star ($\sim 40 M_{\odot}$) collapses to a black hole with an accretion disk and launches a GRB jet (either via neutrino anihilation or via a Blandford-Znajek process).  In this model, it is difficult to produce enough angular momentum to sustain the disk/jet over long timescales (Woosley \& Heger 2006), and hence this model has a shorter-duration central engine.  Collapsars have high accretion rates and are generally associated with more powerful GRBs.  The progenitor star is usually a Wolf-Rayet star which produces strong winds that last $\sim 100,000$ yrs out to kiloparsec distances.  However, it is important to point out that Wolf-Rayet stars have erratic mass loss during last 100-1000 years that may cause CSM to deviate strongly from $1/r^{2}$ profile (see, e.g., Mesler et al. 2012, Margutti et al. 2015).} 

\item{{\bf Helium Merger:} In the helium merger model (Fryer \& Woosley 1998, Zhang \& Fryer 2000), a compact object (neutron star or black hole) spirals into the helium core of its binary companion. These progenitors have high angular momentum and therefore tend to have a long lived central engine.  Moreover, the accretion rates can span a wide range of values, and therefore the power of the burst can range from weak to strong - i.e. this model can accommodate a wide range of emitted energies.  Finally, this model predicts a wind profile medium that is not as strong as in the collapsar case, and extends out to only $\sim 0.01 pc$.  However, this system may experience a common-envelope ejection phase that can produce a dense circumburst medium.}

\item{{\bf White Dwarf-Neutron Star/Black Hole Merger:} This model (e.g., Fryer et al. 1999; Belczynski, Bulik, \& Rudak 2002), in which a white dwarf merges with a neutron star or black hole, has high angular momentum but generally limited fuel and low accretion rates.  Hence, we might expect these bursts to produce weaker, longer duration GRBs.  Because of the lack of stellar wind associated with the system, this model predicts the afterglow occurs in an ISM-like circumburst environment. }

\item{{\bf Neutron Star - Neutron Star/Black Hole Merger:} This model (e.g., Paczynksi 1986; Narayan, Paczynksi \& Piran, 1992), usually associated with the short-duration ($T_{90} < 1s$) class of GRBs (Berger 2014), has limited angular momentum and fuel, but high accretion rates. Therefore, this progenitor naturally produces short, relatively weak GRBs. The GRB afterglow from this model is also expected to occur in an ISM-like environment.}

\item{ {\bf Magnetar:} Magnetars have also been proposed as possible progenitors for GRBs (Duncan \& Thompson 1992; Usov 1992).  In this model, the power from the GRB comes from the spin-down of a highly magnetized neutron star (rather than accretion), and they can in principle occur in a variety of circumstellar environments. Rapidly rotating magnetars (because of their long lifetime compared to a black hole-accretion disk system) are often invoked to explain ultra-long GRBs (e.g. Greiner et al. 2015; see, however, Ioka et al. 2016 who point out that the excessively long spin-down time required to power a SN-like bump make magnetars an unnatural explanation for ultra-long GRBs), and the presence of plateaus in the observed X-ray light curves (Lyons et al. 2010, Yi et al. 2014). However, in magnetar scenarios, it is hard to produce energies above $\sim 10^{52}$ erg for the GRB (and so may not be relevant for the GRBs considered in this paper).} 



\end{itemize}

\subsection{Connecting the Progenitor to GRB Observables}
In what follows, we attempt to connect the burst observables to the progenitor properties discussed above.
\begin{itemize}
\item{ {\bf Prompt Duration:}
  The duration of the gamma-ray emission is not a simple reflection of the duration of the central engine.  The observed duration $T_{90}$ is defined as the time it takes a burst to emit from $5\%$ of its total counts to $95\%$ of its total counts in the detector's (gamma-ray) energy band (Kouveliotou et al. 1993).  A burst with low-level emission over an extended time will therefore have an underestimated $T_{90}$.  Assuming, however, that we are able to detect most of the burst emission and correcting for cosmological time dilation $T_{int} = T_{90}/(1+z)$, the intrinsic duration $T_{int}$ can be affected both by the angular momentum in the accreting system as well as the time the disk is fed (generally thought to be a function of the amount of dense material in accreting system).  However, as mentioned in Gao \& Meszaros 2015, the density profile of the central engine can significantly affect $T_{int}$.  In particular, they showed late internal collisions or refreshed external collisions can extend the value of $T_{int}$ (relative to the active time of the central engine) by a factor of 2 or 3.  Therefore, the so-called intrinsic duration can be significantly affected by the external density profile of the gamma-ray burst.}



\item{ {\bf Radio Loud/Quiet:}
  As discussed in \S 3, there are many parameters that play a role in the brightness of the radio flux and there are a number of possible explanations for why a GRB may not have a radio afterglow.  It may simply be that the afterglow flux has declined rapidly due to the density profile of the circumburst medium.  As discussed briefly in \S 4, for the forward shock emission in an ISM medium, $f_{p}(\nu_{sa}) \propto t^{1/2}$,  $f_{p}(\nu_{m}) \propto t^{0}$; in a wind medium $f_{p}(\nu_{sa}) \propto t^{-1/5}$, $f_{p}(\nu_{m}) \propto t^{-1/2}$.  It may be that the density causes the radio emission to be self-absorbed and therefore strongly suppressed compared to optically thin emission. Finally, the presence (absence) of a radio afterglow could be due to the presence (absence) of reverse shock emission. As discussed in \S 4, a high density or wind medium would lead to a short lived reverse shock that might overlap with the prompt emission and therefore not be detectable as a long-lived radio afterglow.}

\item{ {\bf $E_{iso}-T_{int}$ correlation:}
  If the correlation between $E_{iso}-T_{int}$ in our {\em radio quiet} sample is real (recall, it was found only at the $\sim 3\sigma$ level and observational biases may be playing a role; see \S 3.1), then the progenitors of the radio dark bursts must accommodate this. The correlation may simply be a signature of the length of time a disk is fed, implying a longer GRB duration makes a more energetic burst.  If this is the case, we would expect that the isotropic equivalent peak luminosity is {\em not} correlated with intrinsic duration, since we are suggesting it is the duration of the disk that leads to higher energy (the integral of luminosity over time). Indeed we find this to be the case for a sub-sample of bursts (totalling 17, about half the radio-dark sample) for which data on the peak luminosities is available (from Tsutsui et al. 2013).  A Kendell's $\tau$ test indicates no correlation between $L_{p}$ and $T_{int}$, implying that the total energy is reflective of the duration of the emission in the radio quiet case. If this is indeed the explanation for the correlation seen in the left panel of Figure 4, this implies that the spread in luminosities for the radio quiet bursts is fairly small and offers another clue to their progenitors.}
\end{itemize}


 
\subsection{Other Observables Potentially Tied to the Progenitor}
Below, we briefly summarize other GRB observables that may offer important connections to the progenitor system.
\begin{itemize}
\item{{\bf Plateaus and Flares:}
   The presence of so-called plateaus and flares in the X-ray afterglow light curves of GRBs has also been studied extensively and used to help discern the nature of the GRB emission and connect it to the progenitor (e.g., Swenson \& Roming 2014, Burrows et al. 2005, Fan \& Wei 2005, Zhang et al. 2006, Margutti, et al. 2011, Guidorzi et al. 2015, Greiner et al. 2016). Using the data from Swenson \& Roming 2014 and Yi et al. 2014, we looked for the presence of X-ray flares and plateaus in our radio loud and quiet samples.  The data are sparse, but we found X-ray flares in about half of {\em both} the radio loud and radio quiet samples.  At this point, the numbers are too small to statistically analyze any differences between the flare data in these samples. 

 {\em Interestingly, we found the presence of 4 X-ray plateaus {\bf only in the radio quiet} sample, but none in the radio loud sample}. These numbers are admittedly  small and the presence of plateaus in just four radio quiet GRBs may not be significant.  Expanding this sample with additional broadband follow-up will help determine whether this is a statistical fluctuation or whether there is a true correlation between the presence of X-ray plateaus and radio quiet GRBs.}


\item{{\bf Supernova Associations:}
  Both the radio loud and quiet sample have supernova associations, with 8 Type Ic SNe detections in the radio loud sample and 5 Type Ic SNe detections in the radio quiet sample (see Table 1 of Chandra \& Frail 2012). Given the small number, the rates of detection are statistically similar in both samples.   Most models of the long duration class ($T_{90} > 2s$) of GRBs predict an accompanying supernova (Woosley \& Bloom 2006). However, the models need to account for the fact that there is no He evident in the observed spectra of the SNe associated with GRBs (for a summary of the observations, see Hjorth \& Bloom 2012). At first glance, this may pose a problem for the He merger model of GRBs; however, there are several ways to diminish He detection and the lack He in GRB-SNe is not necessarily a problem for this model (see, for example, Frey, Fryer \& Young 2013 and references therein). In any case, the presence of TypeIc SNe in both samples is an important requirement for the progenitors in both the radio quiet and loud samples.}

\item{{\bf Positions in Host Galaxies}
  We also looked for the positions of these GRBs in their host galaxies. In the radio loud and quiet sample, we found positions/offsets for 11 GRBs and 9 GRBs respectively, using data from Blanchard et al. 2016.  Although the numbers are small, there appears to be no obvious difference between the positions of the radio quiet and radio loud GRBs in their host galaxies. The mean of the normalized host galaxy offset was the same in both samples ($R/R_{h} \approx 1.2$, where $R$ is the measured offset of the GRB from the center of the host galaxy, and $R_{h}$ is the radius of the host galaxy).}
\end{itemize}

  Although the size of these sub-samples (those with flares, plateaus, supernovae, and host offsets) are too small for a robust statistical comparison, the presence of X-ray flares, supernova associations, and measured host galaxy offsets in our radio quiet and radio loud data sets appear similar.  X-ray plateaus appeared in {\em only} the radio quiet sample (albeit in only four GRBs).  The similarities between these additional observables in our radio loud and quiet samples may mean that there is in fact not a fundamental difference between progenitors of these two classes.  However,  it may simply suggest that different progenitors share similar properties in terms of their association with massive stars (and hence SNe and positions in galaxies), but are nonetheless fundamentally different, in terms of the central engine and circumburst environments.

\subsection{Does an Obvious Progenitor Emerge?}
 In the end, what can we say about the progenitors of the radio loud and radio quiet GRBs?  It is important to keep in mind that we have selected only the most energetic bursts ($E_{iso} > 10^{52} erg$) in our analysis, and are therefore selecting for progenitors that can produce these energetic bursts. Keeping this energy selection in mind, both the collapsar and He-merger models are good candidates for the progenitors of our radio dark and bright samples.  Given our most robust result - that radio dark bursts have on average shorter prompt gamma-ray durations - we can ask how these two models accommodate our data.  

 As discussed above, there are many ways to connect the observables (duration and radio flux) to the progenitor, and several interpretations we can offer in the context of the collapsar and He-merger models.  If the radio afterglow is emission from the forward shock of the external blast wave, a natural interpretation is that a dense circumburst medium produces both an observable radio afterglow and a longer duration burst.  In this case, the prompt duration is not just the active time of the central engine, but a reflection of the number and extent of shock events in an extended dense medium, and the radio afterglow is correspondingly brighter in such a medium (assuming it is not self-absorbed).  The collapsar model is a promising candidate to explain the radio bright bursts in this scenario (and can also easily accommodate the on-average higher isotropic emitted energies of the radio bright bursts).

 If the radio afterglow is the result of emission from the reverse shock (as suggested by Mundell et al. 2007, Laskar et al. 2013, Kopac et al. 2015), strong winds associated with a collapsar would cause an early, short-lived reverse shock that would preclude the detection of a long-lived radio afterglow. In this case, we necessarily associate the prompt duration with the (low) amount of angular momentum of the central engine of the collapsar, in order to explain why the radio quiet bursts have on average shorter duration. The He-merger model, with its high angular momentum and moderate winds, would account for the longer-duration GRBs as well as a longer-lived, later-time reverse shock producing the radio afterglow.  A more detailed examination of the progenitor models in the context of these results is the subject of a future publication.

\section{Conclusions}
 We have examined a sample of bright gamma-ray bursts (defined by $E_{iso} > 10^{52} erg$) that have been followed up in the radio band (Chandra \& Frail 2012), and compared the intrinsic properties of bursts with and without radio afterglows.  We find that there is a significant difference in the distributions of intrinsic durations between the two samples. In particular, the radio quiet sample shows significantly shorter prompt burst durations, with an average intrinsic duration more than a factor of two shorter than the radio loud sample.  In addition, we found a mild positive correlation ($\sim 3 \sigma$) between intrinsic duration and isotropic energy for the radio quiet sample, but {\em not} the radio loud sample.

  We suggest these results may offer clues to the progenitors of long GRBs, potentially distinguishing between collapsars and various merger scenarios, although the many parameters that play a role in the determining the radio flux obscure an obvious interpretation. Nonetheless, our results suggest a connection between the prompt gamma-ray emission (often believed to reflect primarily the physics of the inner engine) and the later-time radio afterglow (a reflection of the circumburst environment), implying that both likely depend on the circumburst environment. 

If the radio afterglow emission comes from the forward shock of the external blast wave, a zeroth order interpretation is that a sufficiently dense circumburst medium produces both an observable radio afterglow and a longer duration burst (from more/extended shock events in the surrounding medium).  If the radio afterglow is a result of emission from the reverse shock, the presence of a radio afterglow could suggest a more tenuous ISM-like medium (as opposed to a wind) that would give rise to longer-lived reverse shock emission; in this case, the prompt duration appears to be more directly connected to the amount of angular momentum of the inner engine.  

  A potentially important extension of this is to examine in further detail the multi-wavelength properties of the radio loud and quiet samples. As mentioned in the \S 1, other studies have suggested bimodality of the afterglow emission properties of GRBs in the optical and X-ray, and Gendre et al. (2008) suggest that there could be in fact three populations of radio afterglows based on the properties of their X-ray and optical afterglow emission.  Future work should tie the results in all bands to the properties of the progenitor and circumburst medium.

 We note that only a few GRBs of the short-duration class ($T_{90} < 2s$) have an observed radio afterglow.  Their isotropic-equivalent energies are in general a couple of orders of magnitude lower than the long-duration class of GRBs so it is difficult to extend this analysis (i.e. selecting the brightest bursts to avoid contamination effects) to this class of bursts.  However, it is worth considering the presence/absence of the trends we see in this paper in the context of short GRBs.  We may be able to add significantly to the sample of both long and short GRBs with radio follow-up with the advent of more sensitive radio telescopes. In addition, continued broad band follow-up of GRBs will allow us to get a better handle on whether there truly exist two populations of GRBs and what we might learn about their progenitors from the presence - or lack - of their radio afterglows. 


\acknowledgments{Acknowledgements:}
We are very grateful to Paul Hancock, Bryan Gaensler, and Dale Frail for helpful comments and suggestions on this manuscript. We also gratefully acknowledge the referee for valuable comments that led to an improvement of this manuscript. This work is supported in part by the M. Hildred Blewett Fellowship of the American Physical Society, www.aps.org.  Work at LANL was done under the auspices of the National Nuclear Security Administration of the U.S. Department of Energy at Los Alamos National Laboratory LA-UR-16-27152

\begin{deluxetable*}{lccc}[htbp]
\tablecaption{Radio Bright GRBs}
\tablecolumns{4}
\tablewidth{3.0in}
\tablehead{GRB & $E_{iso} (10^{52}$erg) & $T_{90}$ (s) & $z$}
\startdata
970828 &   29.6   & 147   & 0.96 \\   
980329  &  210  &  58  &  3.9    \\
980703  &  6.9  &  90. &   .96    \\
990123 &   239.    &   100.   &    1.6  \\
990506  &  94.9  &     220    &   1.3  \\
991208  &  11.   &    60   &    .71  \\
991216  &  67.5   &    25.   &    1.02 \\
000131  &  184.   &    110.   &    4.5  \\
000210   &  20.   &    10.   &    .85   \\
000301C  &  4.37   &    10.    &   2.03 \\
000418  &  7.51   &    30.    &   1.12 \\
000911  &  88.0    &   500.    &   1.06\\
000926  &  27.  &     25.   &    2.04  \\
010222  &  133.  &     170.  &     1.48\\
011121 &   4.55   &    105.  &     .36  \\
011211B &   6.3    &   400.  &     2.1   \\
020124 &   30.    &   41.   &    3.2   \\
020405  &  11.0   &    40.   &    .69  \\
020813  &  80.    &   113.   &    1.25 \\
021004  &  3.8    &   50.  &     2.33  \\
030115A &   3.91  &     36.   &    2.5 \\
030226  &  12.0   &    69.   &    1.99 \\
030323  &  3.39   &    20.   &    3.37 \\
030329  &  1.8    &   63.    &   .17   \\
050315  &  5.7  &     96.  &     1.95 \\
050401  &  32.   &    33.   &    2.90 \\
050525A  &  2.04  &     9.0   &    .61\\
050603  &  50.   &    12.   &    2.82 \\
050730  &  9.0    &   157.    &   3.97\\
050820A  &  20.  &     240.  &     2.62\\
050904  &  130.    &   174.   &    6.29\\
050922C  &  3.9   &    5.   &    2.2   \\
051022  &  63.   &    200.   &    0.81\\
051109A  &  2.3   &    37.  &     2.35\\
051111  &   6.0   &    46.   &    1.55\\
060418  &  10.   &    103.   &    1.49\\
061121   & 19.   &    81.   &    1.32 \\
061222A  &  10.3    &   72.   &    2.09\\
070125 &   95.5   &    60.    &   1.55 \\
071003  &  32.4    &   148.  &     1.6 \\
071010B  &  2.6    &   36.   &    .95  \\
070120  &  8.91    &   4.   &    2.15 \\
080319B  &  145   &    125.   &    .94\\
080810 &   53.7    &   108.   &    3.35\\
090313 &   4.57   &    71.    &   3.38 \\
090323  &  410.   &    133.  &     3.57\\
090328  &  10.   &    57.    &   .74   \\
090418  &  25.7   &    56.   &    1.61 \\
090423  &  11.0  &     10.    &   8.26 \\
090618  &  22.1   &    113.  &     .54 \\
090715B   &  23.6  &     265  &     3.  \\
090812 &   44.   &    75.   &    2.45  \\
090902B  &  309.   &   22.0   &   1.88  \\
091020  &  4.56    &   39.   &    1.71 \\
100414A  &   77.9    &   26.   &    1.37\\
100814A   &    5.97    &   175    &   1.44\\
100901A   &    1.78    &   439.   &    1.4\\
100906A   &    13.4    &   114  &     1.73\\
101219B    &   2.96  &     34.  &     .55 \\
\enddata
\tablecomments{Properties of radio bright GRBs used in this analysis. Data taken from Chandra \& Frail, 2012.} 
\label{}
\end{deluxetable*}

\begin{deluxetable*}{lccc}
\tablecaption{Radio Dark GRBs}
\tablecolumns{4}
\tablewidth{3.0in}
\tablehead{GRB & $E_{iso} (10^{52}$erg) & $T_{90}$ (s) & $z$}
\startdata
971214   &    21.0    &    35.  &    3.42   \\ 
990705  &     18.2   &    42.   &    0.84  \\ 
020127   &    3.57  &     26.  &     1.9   \\ 
021211   &    1.1    &   8.   &    1.01   \\ 
030429   &    2.19   &    25.   &    2.66 \\ 
030528   &    3.04    &   84.   &    0.78 \\ 
040924   &    1.10    &   5.   &    0.86  \\ 
041006   &    3.50   &    25.   &    0.72 \\ 
050319   &    4.6    &   153.   &    3.24  \\ 
050408   &    3.44   &    15.   &    1.24 \\ 
050814   &    6.0    &   151.  &     5.3   \\ 
060206   &    4.07   &    8.    &   4.05   \\ 
060210     &  42.   &    255.   &    3.91  \\ 
060522    &   7.00    &   71.  &     5.11  \\ 
060605    &   2.50    &   79.    &   3.77 \\ 
060707     &  6.10   &    66.    &   3.43  \\ 
060908     &  7.00   &    19.  &     1.88 \\
060926    &   1.0    &   8.  &    3.21   \\ 
061126    &   8.0    &   71.   &      1.16 \\ 
061222B    &    8.0   &    40.  &    3.36 \\ 
070306     &  6.0    &   210.  &     1.50 \\ 
070714B     &   1.1     &  3.   &   0.92  \\ 
070721B     &   31.3   &    32.   &    3.63 \\ 
071112C     &  1.95     &  15.  &    0.82  \\ 
080413A    &    13.8    &   46.   &    2.43 \\
080413B     &   1.59   &    8.   &    1.10  \\ 
080603A     &   7.7    &   59.   &    2.69 \\ 
080913    &   6.46   &    8.   &    6.73  \\ 
081118     &  2.8     &  49.   &    2.58  \\ 
081203A     &   34.7    &   223.   &    2.05\\
081222    &   35.4    &   33.    &   2.77  \\ 
090102    &   19.9    &   29.   &    1.55  \\ 
090205    &   2.95    &   9.    &   4.65    \\
090429B     &   5.56   &    5.5   &    9.4 \\
090809     &  1.39    &   8.   &    2.74   \\ 
091127     &  1.41   &    7.  &    0.49   \\ 
110106B     &   3.05    &   25.   &    0.62\\ 
\enddata
\tablecomments{Properties of radio dark GRBs used in this analysis. Data taken from Chandra \& Frail, 2012.}
\label{}
\end{deluxetable*}

\end{document}